\input{aipcheck}

\documentclass[
    ,final            
  ]
  {aipproc}

\layoutstyle{8x11single}

\begin{document}

\title{Non-thermal WIMPs as Dark Radiation}


\author{Farinaldo S. Queiroz}{
  address={Department of Physics and Santa Cruz Institute for Particle Physics University of California, Santa Cruz,
CA 95064, USA}
}

\begin{abstract}
It has been thought that only light species could behave as radiation and account for the dark radiation observed recently by Planck, WMAP9, South Pole and ATACAMA telescopes. In this work we will show that GeV scale WIMPs can plausibly account for the dark radiation as well.  Heavy WIMPs might mimic the effect of a half neutrino species if some of their fraction were produced non-thermally after the thermal freeze-out. In addition, we will show how BBN, CMB and Structure Formation bounds might be circumvented.
 
\end{abstract} 

\maketitle


\section{Introduction}

Among the variety of dark matter candidates, WIMPs stand as one of the most compelling candidates. They are often thought to be GeV-TeV stable particles which are thermally produced in the early Universe and able to plausibly address many direct and indirect detection signals discussed elsewhere \cite{IDsignals,DDsignals}. Recent precise measurements of the angular power spectrum of the cosmic microwave background (CMB) by a variety of telescopes and 
satellites seem to indicate the existence of an extra component of radiation in the early Universe  which is typically parametrized in terms of the number of effective neutrinos species $N_{eff}$. In summary, there is mild evidence for $N_{eff} > 3$ \cite{Neffgt3}. Here we aim to show that GeV-TeV dark matter particles with sub-dominant, non-thermal production might account for the dark radiation observed recently by the Planck Collaboration, ATACAMA and South Pole Telescopes similarly to Refs.\cite{Neffpaper,Farinaldo1,moduli,susyNeff}, as oppose to works where only light species are invoked \cite{Neffmodels}. For concreteness we assume that $1\%$ of the WIMPs are non-thermally produced by the decay of a heavy particle $X^{\prime}$ via $X^{\prime} \rightarrow DM + \gamma$.  Most importantly, any particle physics model that has a WIMP in its spectrum and satisfies the following criteria is able to provide an alternative and interesting solution to this excess of neutrino species:

\begin{itemize}
\item $M_{X^{\prime}}/M_{\rm DM} \geq 4\times 10^5\ \Delta N_{eff}$.

\item ${\it freeze-out\ time} <\ lifetime (X^{\prime})\ < 10^4$ s.

\item Just a small fraction ($\sim 1\%$ or smaller) of the WIMPs should be produced via this non-thermal mechanism.
\end{itemize}

The lifetime may have to be much smaller than $10^4$ s depending on the other particle produced in the final state in addition to the WIMP. For simplicity we restrict ourselves to the pure electromagnetic case (photon), but in principle, this other particle could be anything \cite{hadroniclimits}. Here we will explain from first principles how this mechanism works and in addition we present both a low scale supersymmetry example and a 3-3-1 model where this framework is realized. We will also discuss the constraints imposed by Big Bang Nucleosynthesis, Structure Formation, and the CMB. We begin by deriving the relation between the non-thermal production of WIMPs and the number of effective neutrino species.

\section{WIMPs-Dark Radiation Relation}
\label{darkradiationsec}

To do so, we compute the ratio between an effective neutrino species and dark matter.  Since the Cold Dark Matter (CDM) and neutrino densities are given by $\rho_{DM} = \rho_c \Omega_{DM} a^{-3}$ and $\rho_{\nu} = \rho_c \Omega_{\nu} a_{eq}^{-4} N_{\nu}/3$, at the Matter-Radiation Equality (MRE), the ratio between their energy density is,

\begin{equation}
\frac{\rho_{\nu}}{\rho_{DM}}= \frac{\Omega_{\nu}}{\Omega_{DM}}\ \frac{N_{\nu}}{3}\ \frac{1}{a_{eq}}=\frac{0.69\ \Omega_{\gamma}}{\Omega_{DM}} \frac{N_{\nu}}{3} \frac{1}{a_{eq}},
\label{rhoDMnu}
\end{equation}where $\Omega_{\gamma} \simeq 4.84 \times 10^{-5}$, $\Omega_{DM} \sim 0.227$, $N_{\nu}$ is the number of neutrinos, and $a_{EQ} \simeq 3\times 10^{-4}$ is the scale factor at MRE.

\begin{figure}
  \includegraphics[height=.28\textheight]{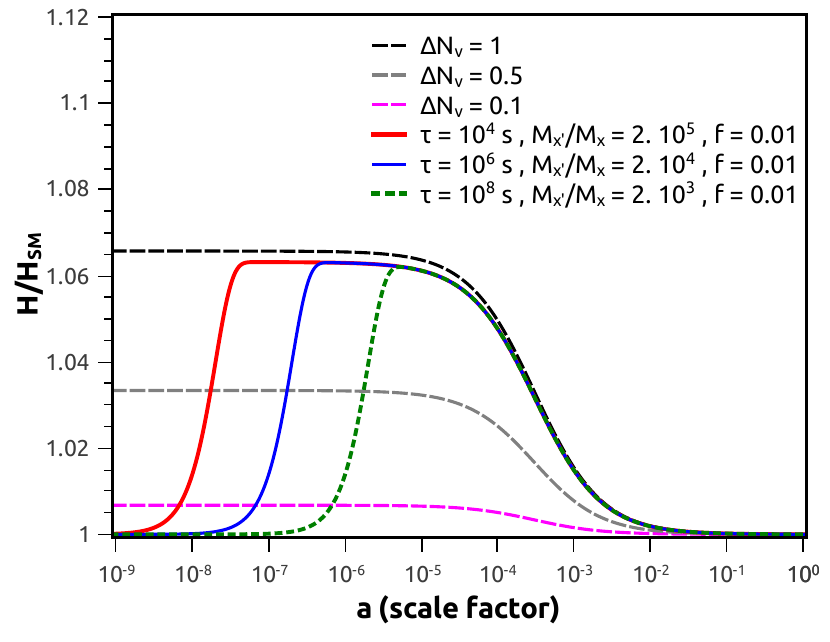}
  \includegraphics[height=.28\textheight]{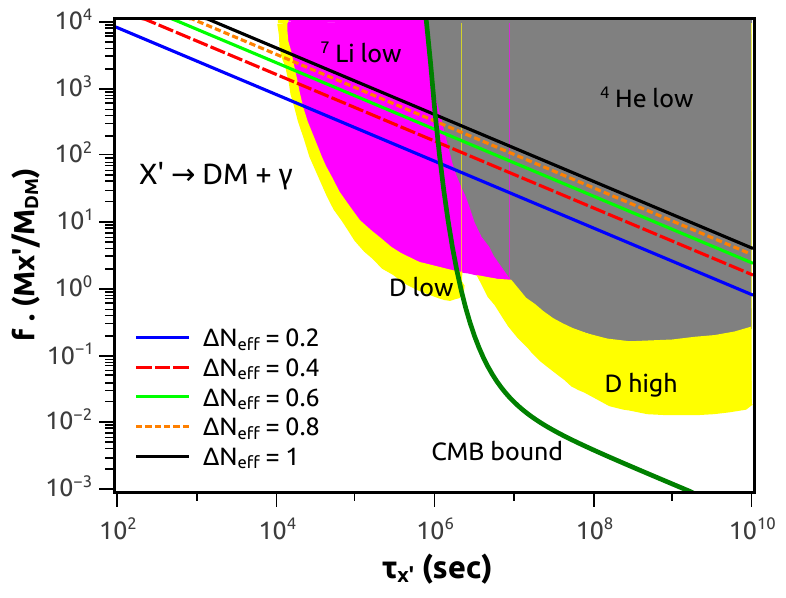}
  \caption{{\it Left}: We exhibit the expansion rate of the Universe for additional neutrino species in dashed lines from top to bottom respectively as $N_{\nu}=3.1 ({\rm pink}),3.5 ({\rm gray}),4 ({\rm black})$. Moreover, we show the expansion rate when we include this non-thermal production of WIMPs that reproduces $N_{eff}=4$ when the mother particle decays with a lifetime of $\tau=10^4{\rm s ( red)},10^6{\rm s (blue)},10^8{\rm s (dashed\ green)}$ \cite{Neffpaper}. 
  {\it Right}: The parameter space defined by the mother particle $X^\prime$ lifetime ($\tau_{X^\prime}$) and the fraction of relativistically produced DM ($f$) times the mother-to-daughter mass ratio $M_{X^\prime}/M_{DM}$, and constraints from BBN and CMB. The shaded regions show BBN bounds on the non-thermal production of DM via the decay $X^{\prime} \rightarrow DM + \gamma$. The green curve represents the CMB bound (regions to the right of the curve are excluded). The diagonal lines corresponding to an excess relativistic degrees of freedom $\Delta N_{eff}=0.2,0.4,0.6,0.8,1$.  We have assumed $M_{X^{\prime}} \gg M_{DM}$. \cite{Farinaldo1}}
\label{figure1}  
\end{figure}

For $N_{\nu}=1$, we thus find that the energy density of one neutrino species is $\sim16\%$ of the CDM density. Hence, if all DM particles had a boost factor of $\gamma_{DM} \simeq 1.16$ at MRE, they would mimic the same effect of an extra neutrino species in the expansion of the Universe at MRE, as can be seen in the left panel of Fig.\ref{figure1} \cite{Neffpaper}. In order to understand how this setup works, we will explain how to relate these two quantities. 

As aforementioned, we are assuming in this work that only a small fraction $f$ of the DM particles in the Universe were produced via the decay $X^{\prime} \rightarrow WIMP + \gamma$. Therefore, using energy and momentum  conservation we get,
\begin{eqnarray}
\label{daneq1}E_{X^{\prime}} &=& E_{\gamma} + M_{DM}\ \gamma_{DM},\\
P_{X^{\prime}} &=& E_{\gamma} + P_{DM}.
\label{daneq2}
\end{eqnarray}where $\gamma_{DM}$ is the boost factor of the dark matter particle.
Assuming that $X^{\prime}$ decays at rest, we get
\begin{equation}
| P_{DM} | = | E_{\gamma} |=  \frac{1}{ 2M_{ X^{\prime} } }(M^2_{X^{\prime}}-M^2_{DM}).
\label{daneq3}
\end{equation}
Substituting Eq.~(\ref{daneq3}) into Eq.~(\ref{daneq1}) and taking $E_{X^{\prime}}=M_{X^{\prime}}$ we find,
\begin{equation}
\gamma_{DM}=  \left( \frac{ M_{X^{\prime}} }{2M_{DM}} + \frac{M_{DM}}{2M_X^{\prime}} \right).
\end{equation}

\noindent Now we have to take into account the expansion of the Universe in the radiation dominated era. Assuming that all decays happen at the lifetime $t=\tau$, we get
\begin{equation}
\gamma_{DM}= 1+ \left( \frac{\tau}{t} \right)^{1/2} \left[\left( \frac{ M_{X^{\prime}} }{2M_{DM}} + \frac{M_{DM}}{2M_X^{\prime}} -1 \right) \right] .
\label{gammaX}
\end{equation}
Hence at MRE ($t=t_{EQ}$) we finally obtain,
\begin{equation}
\gamma_{DM} \simeq 1+ 7.8\times 10^{-4}\left( \frac{\tau}{10^6 s} \right)^{1/2} \left[\left( \frac{ M_{X^{\prime}} }{2M_{DM}} + \frac{M_{DM}}{2M_X^{\prime}} -1 \right) \right] .
\label{gammaX2}
\end{equation}
Notice that Eq.~(\ref{gammaX2}) gives us the boost factor of a DM particle as a function of time and of the mother-to-daughter mass ratio, for the case where all decays happened when the Universe was radiation dominated. This non-thermal production leads to an extra radiation component given by,
\begin{equation}
\rho_{\mbox{extra}}= f \times \rho_{DM} (\gamma_{DM} -1),
\label{rhoextraEq}
\end{equation}where $f$ is the fraction of DM particles non-thermally produced in the decays. This extra radiation coming from the dark sector is called dark radiation and can be parametrized in terms of the number of effective neutrinos by the fact that $\Delta N_{eff} = \rho_{extra}/\rho_{1 \nu} $, where $\rho_{1 \nu}$ is the number density of one neutrino species at the same epoch.  Therefore, using Eq.~(\ref{rhoDMnu}), we find
\begin{equation}
\Delta N_{eff} = \frac{f (\gamma_{DM} -1)}{0.16}.
\label{Neffeq}
\end{equation}
Thus, substituting Eq.~(\ref{gammaX}) into Eq.~(\ref{Neffeq}), we obtain
\begin{eqnarray}
\Delta N_{eff} & \simeq & 4.87 \times 10^{-3}\left( \frac{\tau}{10^6\ s} \right)^{1/2}\nonumber\\ 
                     &   &
  \times \left[\left( \frac{ M_{X^{\prime}} }{2M_{DM}} + \frac{M_{DM}}{2M_X^{\prime}} -1 \right) \right]\times f.
\label{deltaNeff2}
\end{eqnarray}
This equation informs us that if some fraction of the DM particles of the Universe is produced non-thermally through the decay $X^{\prime} \rightarrow DM + \gamma$, this non-thermal production mechanism might mimic a neutrino species for reasonable values of the lifetime and mass ratio ($M_{X^{\prime}}/M_{DM}$) \cite{Neffpaper}. In summary this scenario possesses three free parameters: 
\begin{itemize}
\item[(i)] the $X^\prime$ lifetime for the decay process, 
\item[(ii)] the mass ratio $M_{DM}/M_{X^\prime}$, and 
\item[(iii)] the fraction $f$ of the DM density produced via the decay. 
\end{itemize}

It is important to emphasize that in this framework the majority of the DM particles must still be produced thermally, with only a small fraction being produced non-thermally as we shall discuss further. Now that we have shown how the non-thermal production of DM particles can mimic some degree of additional neutrino species, we will investigate the cosmological bounds which apply to this setup.

\section{COSMOLOGICAL BOUNDS}
\label{sec:cosmo}
\subsection{Structure Formation}

It is a well known fact that all dark matter particles could not have had a large kinetic energy at the matter-radiation equality. DM particles with large kinetic energies would not cluster at sufficiently small scales due to their large free-streaming. Hence, it is critical to check the suppression on the growth of structure caused by the fraction of DM particles which were non-thermally produced in this scenario. At small scales, the matter fluctuations of cold DM particles is governed by a linear equation according to \cite{Ma},
\begin{equation}
\delta (f) \propto  a^{\alpha_{\infty}(f)}.
\label{delta}
\end{equation}where $\alpha_{\infty}(f)$ is the growth rate of the cold DM field at the matter-dominated epoch which is given by \cite{Ma}
\begin{equation}
\alpha_{\infty}(f) = \frac{5}{4}\sqrt{1-\frac{24}{25}f} \simeq 1 - 3/5 f,
\label{alpha}
\end{equation}where $f$ is the fraction of the DM density produced non-thermally in a relativistic state.
Comparing the matter fluctuation given in Eqs.(\ref{delta})-(\ref{alpha}), we can determine the suppression caused by this non-thermal production of WIMPs. To first order, we find

\begin{equation}
g = \frac{\delta (f)}{\delta (f=0)} = a_{EQ}^{-3/5 f} \simeq \exp(-4.9 f),
\label{g}
\end{equation}which is only valid in the matter-dominated regime and for $f\ll 1$. Combined measurements of the amplitude of matter fluctuations in the Universe on scales of $8 h^{-1}$~Mpc from the WMAP9 results \cite{WMAP9} with Lyman-alpha forest data \cite{Lyman} require $g > 0.95$. This bound implies that $f < 0.01$ from Eq.~(\ref{g}). 

In summary, structure formation bounds the fraction of DM particles that can be non-thermally produced in a relativistic state and requires that only a small fraction (less than $1\%$ or so) of the DM in the Universe might have been produced by the decay process being considered here.

\subsection{Big Bang Nucleosynthesis}

As we already pointed out, the lifetime and the energy released by the mother particle $X^\prime$ are also constrained by BBN bounds. The energy released at a given time in the history of the Universe may induce electromagnetic showers that create and/or destroy light elements synthesized in the early universe \cite{BBN}. Thus we should investigate the possible impacts of this scenario on BBN. Given the quite impressive success of BBN we must demand that new physics effects do not drastically alter any of the light element abundances such as those of D, ${}^4$He, ${}^7$Li. The energy of photons created in late decays of $X^{\prime}$ would have been rapidly redistributed through scattering off background photons ($\gamma \gamma_{BG} \rightarrow e^+ e^-$) as well as through inverse Compton scattering ($e\ \gamma \rightarrow e\ \gamma$) \cite{BBN,BBN2,Farinaldo1}. Hence, the bounds we obtain from BBN are, to a good approximation, independent of the initial energy distribution of the injected photons and are only sensitive to the total energy released in the decay process \cite{BBN}.

In order to derive BBN limits on the fraction of relativistic, non-thermally produced DM via the decay $X^{\prime} \rightarrow DM + \gamma$, we need to calculate the total electromagnetic energy released. Let $Y= n/n^{CMB}_{\gamma}$, where $n$ is the number density of particles of a particular species and $n^{CMB}_{\gamma}$ is the number density of CMB photons.  As given by \cite{Neffpaper,Farinaldo1}, we find
\begin{equation}
n^{CMB}_{\gamma}= \frac{2\zeta(2)}{\pi^2} T^3.
\end{equation}
The total electromagnetic energy released from the $X^\prime$ decay is thus $\varepsilon_{EM}= E_{\gamma} Y_{DM}$. If for each  $X^{\prime}$ particle we have the production of a DM particle plus a photon, then $Y_{X^{\prime}}=Y_{\gamma}=Y_{DM,\tau}=Y_{DM,0}$, where $Y_{DM,\tau}$ determines the number density of particles at a time equal to the lifetime of $X^{\prime}$, and $Y_{DM,0}$ is the number density of DM particles today. 

We thus find that the normalized number density of DM particles is given by  \cite{Neffpaper,Farinaldo1},
\begin{equation}
Y_{DM}=\frac{n_{DM}}{n^{CMB}_{\gamma}}= \frac{\Omega_{DM} \rho_c}{M_{DM} n^{CMB}_{\gamma,0}}.
\label{YDM1}
\end{equation}This can be rewritten as,
\begin{equation}
Y_{DM} \simeq 3\cdot10^{-14}\left( \frac{\rm TeV}{M_{DM}} \right) \left( \frac{\Omega_{DM}}{0.227} \right) \left( \frac{f}{0.01}\right).
\label{Yx}
\end{equation}

The factor $f$ showed up in  Eq.~(\ref{Yx}) because we assume here that only a fraction of the DM in the Universe is produced in the decay process, whereas the majority of it is produced non-relativistically by some other mechanism that does not induce any significant energy injection during BBN (for example, via a standard thermal freeze-out process).

Since the photon energy produced in the decay is, 
\begin{equation}
E_{\gamma} =\frac{1}{ 2M_{ X^{\prime} } }(M^2_{X^{\prime}}-M^2_{DM}),
\end{equation}
using Eq.(\ref{Yx}) we find that the total electromagnetic energy released is given by
\begin{eqnarray}
\varepsilon_{EM} & = & 1.5\cdot10^{-11}\ \mbox{GeV}\times \nonumber\\
                 &   &
  \left( \frac{\Omega_{DM}}{0.227} \right) \left( \frac{f}{0.01}\right)  \left( \frac{M_{ X^{\prime} }}{M_{DM}} - \frac{M_{DM}}{M_{X^{\prime}} } \right).
\label{zeta}
\end{eqnarray}

In the limit $M_{X^{\prime}} \gg M_{DM}$ we can straightforwardly connect the total energy release given in Eq.~(\ref{zeta}) with the quantity $f\times\left(M_{ X^{\prime}}/M_{DM}\right)$ as well as with $\Delta N_{eff}$, as given in  Eq.~(\ref{deltaNeff2}). Hence we can translate the results on the total energy released obtained in \cite{BBN}, and express them in terms of the quantity $f\times (M_{ X^{\prime}}/M_{DM})$. We exhibit in Fig.~\ref{figure1} the shaded regions ruled out by BBN using the ``baryometer'' parameter $\eta=n_b/n_{\gamma}=6\times 10^{-10}$. The diagonal lines represent combinations of the quantity $f\times M_{ X^{\prime}}/M_{DM}$ and $\tau_{X^\prime}$ producing the $\Delta N_{eff}$ as in the labels.

We conclude from Fig.~\ref{figure1} that the BBN bounds are weaker for early decays because at early times the
universe is hot enough and the initial photon spectrum is rapidly thermalized, leaving just a few high-energy photons able to modify the light element abundances. Although for lifetimes longer than $10^4$~s, BBN excludes most of the relevant parameter space.  

\subsection{Cosmic Microwave Background}

Similarly to the BBN constraints, CMB bounds depend mostly only on the total energy released in the decay process.
The key effect of the additional energy injection in the form of photons is related to spectral distortions caused in the CMB black-body spectrum \cite{CMBbound}. For early times ($t < 10^3$~s) the processes of bremsstrahlung, i.e. $eX \rightarrow eX\gamma$ (where X is an ion), Compton scattering and double Compton scattering $e \gamma \rightarrow e \gamma \gamma$  quickly thermalize the injected photon energy \cite{CMBbound}. On the other hand, for $t > 10^3$~s, the bremsstrahlung and double Compton processes become inefficient, and the photon spectrum relaxes to a Bose-Einstein distribution with a chemical potential ($\mu$) different from zero. Limits on $\mu$ are used to constrain this additional energy injection and consequently bound the set $fM_{X^{\prime}}/M_DM$ and the lifetime as discussed in Ref.\cite{Neffpaper,Farinaldo1}. The current limit implies that $\mu < 9 \times 10^{-5}$ \cite{cmb1, cmb2}. In Fig.~\ref{figure1} we plotted the CMB bound after converting this upper limit on the chemical potential into a bound on $f M_{X^{\prime}}/M_DM$ for a given lifetime. Notice that CMB constraint becomes only relevant for lifetimes longer than $\sim 10^6$~s. 

We have discussed the dark radiation setup so far and now it is time to present some realistic models where this scenario is plausibly realized while simultaneously obeying the cosmological limits aforementioned.

\section{Supersymmetric framework}

Neutralinos are the mass eigenstates resulting from a mixture of neutral B-ino, W-ino, and Higgs-inos. Here we will assume it to be a pure Bino. In low scale supersymmetry the neutralino might be the next-to-lightest supersymemtric particle with the lightest supersymmetric particle being the gravitino. Thus Binos decay into a gravitino-photon final state via the interaction Lagrangian term \cite{Jfeng1},
\begin{equation}
\mathcal{L}= \frac{-i}{8\pi M_{\star}} \tilde{G_{\mu}}[\gamma^{\nu},\gamma^{\rho}]\gamma^{\mu}\tilde{B^{\mu}}F_{\nu \rho},
\label{eqgravitino}
\end{equation}where $M_{\star}=2.4\times 10^{18}$~GeV is the reduced Planck mass. 
Because we are in the regime of low-scale supersymmetry breaking $M_{\tilde{B}} \gg M_{\tilde{G}}$, which is exactly the limit needed to realize the dark radiation setup we are focused on. In this case, from Eq.~(\ref{eqgravitino}) we find a neutralino lifetime of
\begin{equation}\label{eq:binolifetime}
\tau (\tilde{B} \rightarrow \gamma  \tilde{G}) \simeq 750\ {\rm s}\left(\frac{M_{\tilde G}}{1\ \rm keV}\right)^2\left(\frac{1\ {\rm GeV}}{M_{\tilde B}}\right)^5.
\end{equation}
Solving Eq.(\ref{eq:binolifetime}) for $M_{\tilde B}$ and substituting into Eq.(\ref{deltaNeff2}) we find a lower
limit on the DM mass,
\begin{figure}[!t]
\centering
\includegraphics[height=.27\textheight]{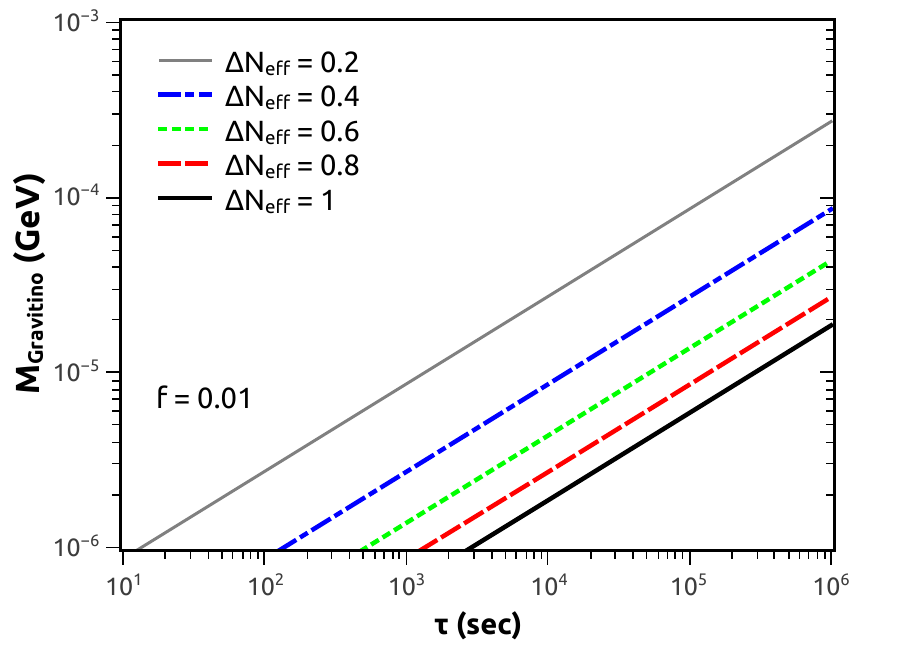}
\includegraphics[height=.27\textheight]{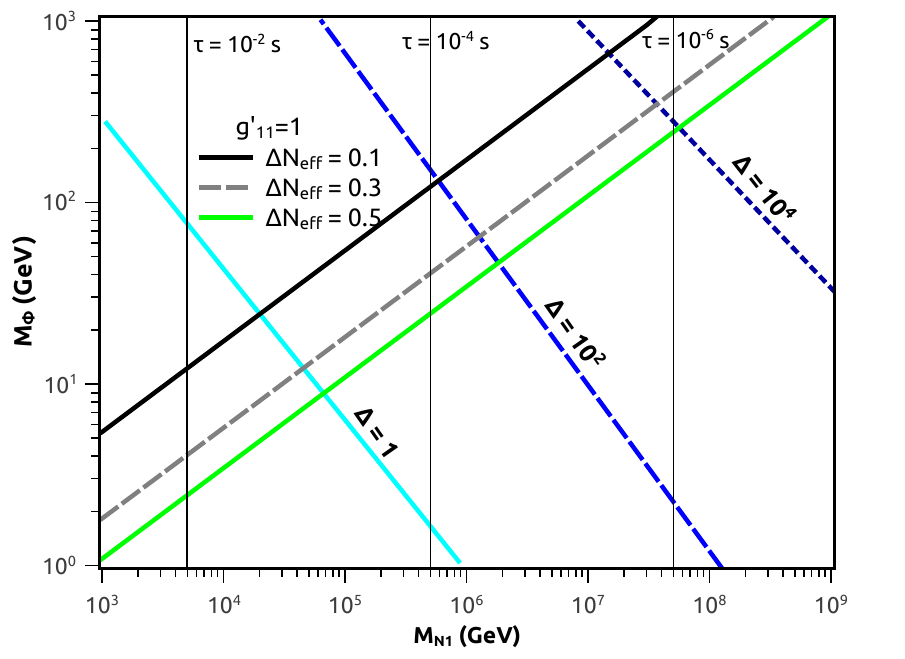}
\caption{{\it Left:}Lower limit on the gravitino mass, for a supersymmetric model where a fraction $f$ of gravitinos are non-thermally produced by the decay of pure Binos. The relevant lifetime is derived from Eq.~(\ref{eqgravitino}). We may conclude that a $\sim 10$~KeV Gravitino might reproduce $\Delta N_{eff} \sim 0.1$. {\it Right:} Bound on the WIMP ($\phi$) mass of the 3-3-1LHN model. Notice that a $100-1000$~GeV is perfectly able to reproduce $\Delta N_{eff} \sim 0.1$. We have used $f=0.01$ in both plots.}
\label{figravitino}
\end{figure}
\begin{equation}
M_{G } < (4\ {\rm MeV}) \left( \frac{\tau}{10^4}\right)^{1/2} \left( \frac{f}{\Delta N_{eff}}\right)^{5/3}.
\end{equation}

The lower bound is found because we imposing the lifetime to be shorter than $10^4$~s to obey BBN limits. Setting $f=0.01$ and $\tau=10^4$ s, we show in Fig.~\ref{figravitino} that a gravitino with mass in the 2 to 20 keV range mimics the effect of an extra neutrino species while still obeying cosmological bounds. Notice that for a gluino mass of $\sim 1$~TeV a reheating temperature $T_R$ close to the electro-weak scale is required to prevent over-closing the universe, since the thermal production of gravitinos in the early universe  \cite{Bolz:2000fu} implies
\begin{equation}
\frac{T_R}{100\ {\rm GeV}}\simeq \left(\frac{\Omega_{\tilde G}h^2}{0.2}\right)\left(\frac{1\ {\rm keV}}{M{\tilde G}}\right).
\end{equation}
Such a low reheating temperature would rule out certain scenarios for the production of the baryon asymmetry in the universe, such as Leptogenesis, but is in general not phenomenologically implausible. 

Finally, we note that the constraints on the lifetime and Eq.~(\ref{eq:binolifetime}) imply a lower limit on the bino mass, which must be larger than about 1 GeV.  This is, of course, perfectly compatible with the supersymmetric models relevant here.

\section{3-3-1 Model}

3-3-1 models  refer to  electroweak extensions of the Standard Model gauge group based on the enlarged gauge group $ SU(3)_c\otimes SU(3)_L\otimes U(1)_N$. Models based on this 3-3-1 gauge symmetry \cite{331versions} potentially address important theoretical and phenomenological questions which remain unexplained within the SM, such as the number of particle generations \cite{families}, the possible Higgs to diphoton excess \cite{331higgs} and have a rich phenomenology which includes new scalars and gauge bosons, as extensively explored in the literature \cite{331feno}. In particular, more recent work based on this symmetry have been proposed to additionally address debatable direct and indirect detection signals of WIMPs in our galaxy \cite{331DM_1,331DM_2,331DM_3,331DM_4,331DMversions}\footnote{Models based on the 3-4-1 gauge symmetry which embeds the 3-3-1 might also have good dark matter candidates (see Refs.\cite{341}).}. For these and many other reasons, 3-3-1 models stand as compelling alternatives to the SM. Here we will focus on a version that has heavy neutrinos ($N_{1R}$) and a stable scalar (the WIMP of the model) in its spectrum known as 331LHN described in more details in Ref.~\cite{331DM_4}. In this model the so called WIMP miracle is realized and the right abundance is easily achieved. Although the heavy neutrinos can be long lived, in the sense that they may decay after the WIMP ($\phi$) freeze-out, some non-thermal production of $\phi$ will occur. In supersymmetric models, the supersymetric particles will always decay into the lightest stable particle due to the R-parity symmetry.  Similarly, because of a global symmetry (G), described in \cite{331DM_4}, the 331 particles will decay into $\phi$ as well, inducing a non-thermal production of dark matter when the decays happen after the WIMP freezes-out. For simplicity we will tune only one heavy neutrino ($N_1R$) to be long lived, but a more general approach could be straightforwardly investigated \cite{331DM_4}. In summary, this model will have dominant thermal production as in the standard WIMP paradigm, but an additional a sub-dominant one arises. This non-thermal production component is crucial in our setup because it generates some degree of dark radiation as shown earlier. Nevertheless, the lifetime of the mother particle $N_{1R}$ has to be shorter than $10^4$~s due to BBN constraints. The most important parameters which control the lifetime of this neutrino are the scale of symmetry breaking of this model, its mass and the Yukawa coupling, $g^{\prime}_{11}$. From our perspective, the relevant decay mode is $N_{1R}\rightarrow WIMP + \nu_e$ which has the following lifetime,
\begin{equation}
\tau \simeq  \left(5\cdot10^{-5}{\rm s}\right)\left(\frac{10^{-3}}{g^{\prime}_{11}}\right)^2 \left(\frac{v_{\chi^{\prime}}}{10^3\ \mbox{GeV}}\right)^2\left( \frac{10^{12}\ \mbox{GeV}}{M}\right) .
\label{lifetime}
\end{equation}where $\lambda = g^{\prime}_{11}v/v_{\chi^{\prime}}$ and $M$ is the see-saw scale. 
From Eq.(\ref{lifetime}) we notice that there will be a very wide range of Yukawa couplings ($g^{\prime}_{11}$) that produces a lifetime allowed by BBN ($\tau \leq 10^4$~s) and with decays that happen after the WIMP freeze-out ($\tau \simeq 10^{-8}$~s for a $100$~GeV WIMP). However, when we try to reproduce $N_{eff} \simeq 3.5$ and simultaneously obey BBN, CMB and structure formation bounds, the parameter space which satisfies all these criterias is rather reduced. 

The parameter space in this scenario includes the mass of the daughter particle $M_{\rm wimp}$, the mother particle mass ($M_{N1}$) and the coupling constant $g_{11}^{\prime}$, which sets the thermal relic densities. In Fig.~\ref{figravitino} we present our results in the mother-daughter mass parameter space for a relatively large  coupling, $g_{11}^\prime=1$. Notice we find a line across the parameter space where all of the constraints are satisfied, and where $\Delta N_{\rm eff}=0.1$ for WIMP masses in the range between a few GeV and a few tens of GeV, and for $N_1$ masses between 10 and 100 TeV. Larger $N_1$ masses require increasingly larger entropy suppression factors $\Delta$, and larger WIMP masses to obtain the desired enhancement to $\Delta N_{\rm eff}$. An entropy injection episode should be invoked for small values of $g_{11}^{\prime}$ because the abundance of the mother particle is typically too large to only produce $1\%$ of the WIMP density as required by structure formation. Therefore we postulate that an entropy injection occurred between the relatively high temperature at which the $N_1$ froze out and the time of decay (the latter is indicated by vertical lines in the figures) causing a dilution factor $\Delta$ in the abundance. In other words, the standard thermal relic density $\Omega_{N1}\to\Omega_{N1}/\Delta$ as a result of the larger entropy density. $\Delta=1$ reproduces the standard cosmological model. It is worth noticing in Fig.~\ref{figravitino} that larger masses are possible if one invokes larger dilution factors.

\section{Conclusions}

Recent measurements of the cosmic microwave background radiation point to a mild evidence for dark radiation, i.e, an excess of relativistic degrees of freedom in the early universe. In this work we have shown that if a fraction of the dark matter particles of the Universe had been produced from the decay of a heavy particle after their freeze-out, this non-thermal production will induce some degree of dark radiation. Indeed, these non-thermal WIMPs would be able to mimic the effect of one neutrino species in the early universe. Furthermore, we have shown that this mechanism must obey strong BBN, CMB and Structure Formation constraints. Nevertheless, they can be circumvented if the lifetime of the mother particle is shorter than $10^4$~s and just a fraction ($\sim 1\%$) of the DM particles had been created non-thermally. Lastly, we presented a low scale supersymmetry framework and 3-3-1 model where such extra radiation arises from a sub-dominant production of DM particles from the decay of a heavy particle while evading BBN, CMB and Structure Formation bounds. In particular, we show that a $\sim 10$~KeV Gravitino, in low scale supersymmetry, and a $\sim 10$~GeV scalar, in the 3-3-1LHN model, are both plausibly able to reproduce $\Delta N_{\rm eff}\simeq 0.5$.

\begin{theacknowledgments}
We are grateful to Stefano Profumo, Chris Kelso, Carlos Pires, Paulo Rodrigues for their collaboration. We thank the organizers of PPC2013 for a well organized conference. This work is partly supported by the Department of Energy under contract DE-FG02-04ER41286 (SP), and by the Brazilian National Counsel for Technological and Scientific Development (CNPq) (FQ,CP,PR).
  
\end{theacknowledgments}



\bibliographystyle{aipproc}   

\bibliography{sample}

\IfFileExists{\jobname.bbl}{}
 {\typeout{}
  \typeout{******************************************}
  \typeout{** Please run "bibtex \jobname" to optain}
  \typeout{** the bibliography and then re-run LaTeX}
  \typeout{** twice to fix the references!}
  \typeout{******************************************}
  \typeout{}
 }

\end{document}